\documentclass[10pt,letterpaper]{article}
\usepackage{opex3}
\usepackage{amssymb}
\usepackage{cite}

\begin{document}

\title{Mathematical model of thermal shields for long-term stability optical resonators}
\author{Josep Sanjuan,$^{1}$ Norman G\"urlebeck$^{2}$ and Claus Braxmaier$^{1,2}$}
\address{$^{1}$German Aerospace Center (DLR), Robert-Hooke-Str. 7, Bremen 28359, Germany}
\address{$^{2}$Center of Applied Space Technology and Microgravity (ZARM), University of Bremen, Am Fallturm 1, 28359 Bremen, Germany}

\email{$^*$jose.sanjuan@dlr.de} %% email address is required



\begin{abstract}
Modern experiments aiming at tests of fundamental physics, like measuring gravitational waves or 
testing Lorentz Invariance with unprecedented accuracy, require thermal environments that are 
highly stable over long times. To achieve such a stability, the experiment including typically an 
optical resonator is nested in a thermal enclosure, which passively attenuates external 
temperature fluctuations to acceptable levels. These thermal shields are usually designed using 
tedious numerical simulations or with simple analytical models. In this paper, we propose an 
accurate analytical method to estimate the performance of passive thermal shields in the 
frequency domain, which allows for fast evaluation and optimization. The model analysis has also
unveil interesting properties of the shields, such as dips in the transfer function for some frequencies 
under certain combinations of materials and geometries. We validate the results by 
comparing them to numerical simulations performed with commercial software based on finite element methods.
\end{abstract}

\ocis{(120.2230) Fabry-Perot; (120.6810) Thermal effects; (120.3930) Metrological instrumentation.} 
% REPLACE WITH CORRECT OCIS CODES FOR YOUR ARTICLE, MINIMUM OF TWO; 

%%%%%%%%%%%%%%%%%%%%%% References %%%%%%%%%%%%%%%%%%%%%%%%%

%%%%%%%%%%%%%%%%%%%%%%%%%%  body  %%%%%%%%%%%%%%%%%%%%%%%%%%

\section{Introduction}
The ever increasing accuracy of current and future fundamental 
physics experiments on ground and in space demands very stable 
and controlled environments to meliorate spurious effects such 
as temperature fluctuations. In this paper, we present a method 
that allows the rapid assessment of the performance of passive 
thermal shields in the frequency domain. This is done with analytical models 
instead of time consuming and precision limited 
numerical simulations. The frequency domain is of special interest 
since stability requirements are typically given in such domain.
%Later on, the design can be fine tuned 
%by simulations, where all sort of details can be included.

One of the key applications of this method is the design of 
thermal isolations for optical resonators (ORs), which are 
used as frequency references in highly sensitive interferometers, 
in experiments testing fundamental physics, 
and as optical frequency standards, see e.g.~\cite{0264-9381-29-12-124016, 0264-9381-26-15-153001, Mueller1, heinzel.grace,GRACEFO}, 
\cite{PhysRevLett.88.010401,PhysRevLett.91.020401,PhysRevD.80.105011,PhysRevD.81.022003, 2012arXiv1203.3914L, 6702260,0264-9381-18-13-312,RevModPhys.83.11,
0034-4885-77-6-062901}, and \cite{PhysRevA.77.033847,Argence:12,Amaira_2013,:/content/aip/journal/rsi/85/11/10.1063/1.4898334}, respectively. 
ORs consist of a spacer and two high-reflectivity mirrors. 
The distance between the mirrors defines the resonance frequency, which
is directly linked to the length of the spacer. Consequently,  
temperature fluctuations result in frequency fluctuations, i.e.,
\begin{equation}
 \frac{\delta \nu}{\nu} = \frac{\delta \ell}{\ell} = \alpha\delta T,
\end{equation}
where $\nu$ is the optical frequency, $\ell$ is the spacer length, $\alpha$ is the spacer coefficient 
of thermal expansion (CTE) and $T$ is its temperature. Fluctuations of the variables 
are expressed as $\delta$. The spacers of ORs are made of materials with very low CTEs, e.g.\ 
ultra low expansion glass (ULE) or Zerodur with CTEs about $10^{-8}$\,K$^{-1}$ 
at room temperature~ \cite{1996SPIE.2857...58E,Birch:88}. In addition, the ORs are often driven at temperatures close 
to their CTE zero crossing (or at the CTE minimum)~\cite{PhysRevA.77.033847,PhysRevA.77.053809}. 
In case these temperatures cannot be achieved, different techniques have been 
developed to tune the temperature of the zero crossing~\cite{0022-3727-28-9-008, Legero:10}. Nevertheless, the 
accuracy of current optical experiments is already so high that these low CTEs do not 
suffice and temperature fluctuations need to be strongly attenuated. 
The damping of temperature fluctuations at frequencies in the milli-Hertz and sub-milli-Hertz regime  
by several orders of magnitude demands a particularly careful thermal shield design. This low frequency range 
is crucial in different missions and experiments since the expected science signals are in the same frequency band.
This is especially important for the space-based gravitational wave detector eLISA~\cite{0264-9381-29-12-124016} and
foreseen missions testing the fundamental principles of general relativity such
as the Lorentz Invariance with Michelson-Morley experiments
and a Kennedy-Thorndike experiment like STAR (SpaceTime Asymmetry
Research~\cite{2012arXiv1203.3914L}), mini-STAR, and BOOST (BOOst Symmetry Test~\cite{6702260}), for earlier Kennedy-Thorndike
experiments see, e.g., \cite{PhysRevLett.88.010401,PhysRevD.81.022003}. 

The thermal shield design is driven by the required temperature stability of 
the OR, i.e., the required frequency stability for a given spacer CTE, and 
the temperature variations at the outer most thermal shield layer. 
For eLISA the required temperature stability (if a cavity is used for 
the laser pre-stabilization) is ~$\mu$K\,Hz$^{-1/2}$ in the mili-Hertz band. 
The thermal environment in eLISA will be extremely stable~\cite{:/content/aip/proceeding/aipcp/10.1063/1.2405044} and 
the required attenuation will not be very demanding. However, the main problem 
is for laboratory-based demonstration experiments~\cite{0264-9381-26-15-153001,PhysRevD.87.102003} where 
$\mu$K\,Hz$^{-1/2}$ temperature stability in the sub-milli-Hertz range is needed.
For the Kennedy Thorndike experiments the allowed temperature 
fluctuations are typically $\lesssim$0.1\,$\mu$K at the orbital period ($\sim$90\,minutes) 
and the expected stability in the satellites around 1\,K
and $\sim$10\,mK can be achieved in the payload by active temperature control. This poses a
stringent requirement on the thermal shields: about five orders of magnitude attenuation 
at $\sim$0.2\,mHz. For the Gravity Recovery and Climate Experiment 
Follow-On (GRACE-FO) interferometer similar figures are required~\cite{GRACEFO}. 
The thermal shield performance is usually estimated with
the help of tedious numerical simulations~\cite{:/content/aip/proceeding/aipcp/10.1063/1.2405044,Argence:12} 
or with rather simplified analytical models. However, a method to estimate its transfer function yielding results 
close to the numerical simulations is to our best knowledge still lacking. In 
this paper, we provide such a tool, using thermal shields for optical resonators as an example.

The paper is organized as follows: First the general mathematical model is
introduced, which is applied to the cases of spherical and
cylindrically shaped shields in Sec.~\ref{sec.4}. These results are compared with numerical simulations 
based on finite element methods (FEM) in Sec.~\ref{sec.5}. Sec.~\ref{conclusion} summarizes our conclusions. Supplementary 
models and mathematical details are provided in the appendix.

\section{Mathematical model of the thermal shields \label{sec.3}}
The basic assumption of the thermal shields' mathematical model is that 
heat is transferred only by radiation and conduction.
In Sec.~\ref{sec.3.1}, the analytical transfer function of the thermal shields is derived, when the latter 
is negligible. In Sec.~\ref{sec.3.2}, the conductive links between the shields are included in the model, too.

\subsection{Radiative heat transfer \label{sec.3.1}}
The radiative heat transfer between two gray bodies (from body $i$ to body $j$) is defined as~\cite{Incropera}:
\begin{equation}\label{eq.2.pep}
 \dot{q}_{j}(t) = \frac{\sigma A_{j} [T_{i}^{4}(t) - T_{j}^{4}(t)]}{\beta_{ij}},
\end{equation}
where $\sigma$ is the Stefan-Boltzmann constant ($=5.67\times10^{-8}\,{\rm W}\,{\rm m}^{-2}\,{\rm K}^{-4}$), 
$A_{j}$ is the area, $T_{j}$ is the temperature in Kelvin of the respective layers and $\beta_{ij}$ is 
a term including the view factors and emissivities, $\varepsilon$ ---see Sec.~\ref{sec.4}. 
The temperature change of the layer $j$ due to the heat exchange given by Eq.\ (\ref{eq.2.pep}) is
\begin{equation}\label{eq.3.pep}
\dot{q}_{j}(t) = m_{j}c_{j}\dot{T}_{j}(t),
\end{equation}
where $m_{j}$ and $c_{j}$ are the mass and the specific heat of the layer $j$, respectively. We 
assume that the thermal shields have high thermal conductivity and are relatively thin. Consequently, 
the temperature distribution in the shields homogenizes rapidly even in the cases, where 
point heat sources are present ---see Sec.~\ref{sec.4}. Hence, we do not consider a spatial temperature distribution. Combining 
Eqs.\ (\ref{eq.2.pep}) and (\ref{eq.3.pep}) results in the differential equation
\begin{equation}\label{eq.5.pep}
 \frac{4\sigma A_{j}T_{0}^{3}}{\beta_{ij}}[T_{i}(t)-T_{j}(t)]=m_{j}c_{j}\dot{T}_{j}(t)
\end{equation}
that has been linearized in order to find an analytical solution. $T_{0}$ is the average temperature of the bodies. 
Typically, temperature changes during the experiments are small compared to $T_0$ and, thus, the linearization is justified.

The transfer function of the thermal shield (from layer $i$ to $j$) is obtained after applying the Fourier 
transform to Eq.\ (\ref{eq.5.pep}) and taking the ratio between both temperatures:
\begin{equation}\label{eq.6.pep}
 \widetilde{H}_{ij}(\omega)=\frac{\widetilde{T}_{j}(\omega)}{\widetilde{T}_{i}(\omega)}
 =\frac{1}{1+\frac{m_{j}c_{j}\beta_{ij}}{4\sigma A_{j}T_{0}^{3}} i\omega },
\end{equation}
which corresponds to a first-order low-pass filter with a cut-off angular frequency of
\begin{equation}\label{eq.7.pep}
 \omega_{\rm c}= \frac{4\sigma  A_{j} T_{0}^{3}}{m_{j} c_{j} \beta_{ij}}.
\end{equation}
Analogously to the electrical case, the cut-off angular frequency is split in a thermal resistance (in units of K\,W$^{-1}$) 
and a thermal capacitance (in units of J\,K$^{-1}$): 
\begin{eqnarray}	
 \theta_{ij} &=& \frac{\beta_{ij}}{4\sigma A_{j}T_{0}^{3}} \label{eq.8.pep},\\
 C_{j} &=& m_{j}c_{j}  \label{eq.9.pep}
\end{eqnarray}
and the product is the time constant of the system, $\tau_{ij}=\theta_{ij}C_{j}=\omega_{\rm c}^{-1}$.

Once the transfer function of one shield (formed by 2 layers) has been determined the behavior 
of $N$ concentric thermal shields ($N+1$ layers) can be calculated. In order to obtain a compact expression, 
it is assumed that all the layers are exactly the same (in material and in size), i.e., 
they all have the same time constant $\tau_{ij}=\tau$. The validity of this assumption is discussed in Sec.~\ref{sec.4}, where 
we show that it holds for thin layers that are sufficiently close to each other. In Appendix~A, the 
transfer function for $N$ layers is derived and the result is 
\begin{eqnarray}\label{eq.10.pep}
 \widetilde{H}(\omega)&=&\frac{1}{1+\sum\limits_{k=1}^{N}\frac{1}{(2k)!}\frac{(N+k)!}{(N-k)!}(i \omega\tau)^{k}}  \nonumber \\
 &=&\left(1+\frac{1}{4} i \omega  \tau \right)^{1/2} \sec \left[(2 N+1) \csc ^{-1}\left(\frac{1+i}{\sqrt{\omega \tau/2 }}\right)\right]. \nonumber \\
\end{eqnarray}

\begin{figure}[h!]
 \begin{center}
    \includegraphics[width=0.8\linewidth]{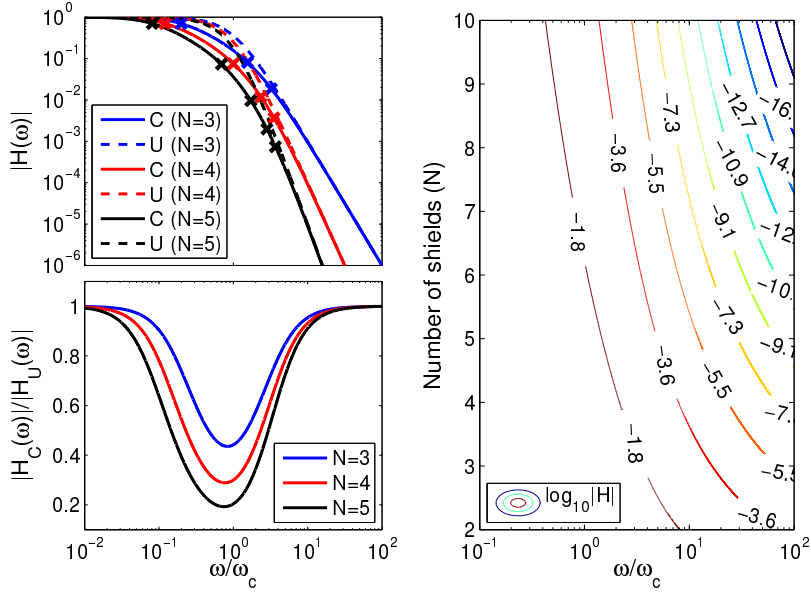}
 \end{center}
\caption{Left: transfer functions using the coupled case Eq.\ (\ref{eq.10.pep}) (solid lines) and the uncoupled case
Eq.\ (\ref{eq:uncoupled}) (dashed lines) for $N=3$, 4 and 5. The cut-off angular 
frequency of each layer is $\omega_{\rm c}$ (cf.\ Eq.\ \ref{eq.7.pep}). Notice that  
some of the poles of the exact solution (cross marks) are at lower frequencies than $\omega_{\rm c}$ (C: coupled filters. U: 
uncoupled filters), which improves the damping for frequencies around $\omega_{\rm c}$. 
For $\omega\gtrsim10\omega_{\rm c}$ the results are the same. The bottom plot shows the 
ratio between the coupled and uncoupled solutions. Right: $|\widetilde{H}(\omega)|$ as a function of the frequency 
and the number of layers (in logarithmic scale). \label{fig.2}}
\end{figure}

This is depicted in Fig.\ \ref{fig.2} for different $N$. The 
left panel shows, how the layers interact with each 
other causing some of the poles of the transfer function, 
cf.\ Eq.\ (\ref{eq.10.pep}), of the system to appear at 
lower frequencies than the poles of each individual layer. 
For this reason, the system exhibits a stronger temperature 
damping at frequencies around the cut-off angular frequency 
than if the layers are assumed uncoupled, i.e., 
\begin{equation}\label{eq:uncoupled}
\widetilde{H}(\omega)=(1+i \omega\tau)^{-N}.
\end{equation} 
However, 
for $\omega\gtrsim10\omega_{\rm c}$ the response of both systems (coupled or uncoupled) is 
the same and Eq.\ (\ref{eq:uncoupled}) can be used instead of Eq.\ (\ref{eq.10.pep}).

\begin{figure}[h!]
 \begin{center}
  \includegraphics[width=0.7\linewidth]{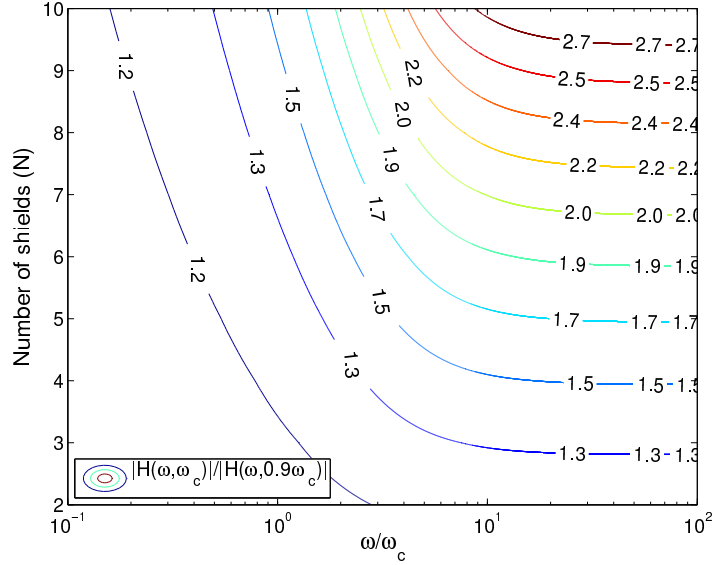}
 \end{center}
\caption{Ratio between $|H(\omega,\omega_{\rm c})|$ and $|H(\omega,0.9\omega_{\rm c})|$ as a 
function of $N$ and angular frequency. The relative error of the cut-off angular frequency is $-$10\%. 
The errors in the transfer function are tolerable even for large $N$ if the 
relative error in $\omega_{\rm c}$ is kept smaller than $-$10\%.
\label{fig.2b}
}
\end{figure}

In order to calculate the insulator transfer function, cf. Appendix~A, the cut-off angular frequency 
needs to be well known to avoid errors in $|\widetilde{H}(\omega)|$. The 
errors in the cut-off frequency are due to the assumption that all the time 
constants of the shields, $\tau$, are the same. The ratio between the transfer 
function considering a $-$10\% relative error in $\omega_{\rm c}$ and the 
actual transfer function is shown in Fig.~\ref{fig.2b}. The errors only become significant 
when large number of layers are used and for $\omega\gg\omega_{\rm c}$. 
%A 10\% error is somewhat a conservative value since typically larger 
%errors are considered acceptable since the different 
%practical implementations and constructions of the shields will  
%likely cause more uncertainties. 
This will be used in Sec.~\ref{sec.4} to asses 
the validity of the assumption that $\tau_{ij}=\tau$. 

For completeness, we present here the transfer function in the case when the 
OR is included. The OR acts as an extra low-pass filter. However, its 
thermal resistance, $\theta_{\rm OR}$, and thermal capacitance, $C_{\rm OR}$, can be, in general, 
different from the thermal shield one, i.e., $\tau\neq\tau_{\rm OR}$. The transfer 
function is ---see Appendix~A for details,
\begin{eqnarray}
 \widetilde{H}(\omega)=&\left[ 1 + \frac{\tau_{\rm OR}}{\tau}(i\omega\tau)^{N+1}+ 
 \sum_{k=1}^{N}\left[\frac{1}{(2k-1)!}\frac{(N+k-1)!}{(N-k)!}\frac{C_{\rm OR}}{C}+\right. \right. \nonumber\\
 &\left.\left. \frac{1}{(2k-2)!}\frac{(N+k-1)!}{(N-k+1)!}\frac{\tau_{\rm OR}}{\tau}+ 
 \frac{1}{(2k)!}\frac{(N+k)!}{(N-k)!}\right](i\omega \tau)^{k}\right]^{-1},\nonumber\\ 
\label{eq.22}
\end{eqnarray}
where $\tau_{\rm OR}=\theta_{\rm OR}C_{\rm OR}$. If $\tau_{\rm OR}\ll\tau$ and $C_{\rm OR}\ll C$, Eq.~(\ref{eq.10.pep}) is 
recovered and the effect of the resonator is negligible. Here $N$ is the number of shields without including the 
resonator. However, in some cases $C_{\rm OR}\geq C$ since optical resonators 
are typically bulky bodies made of Zerodur or ULE. The thermal resistance (between the last layer of the thermal shield and the resonator)
and thermal capacitance of the resonator are:
\begin{eqnarray}
 \theta_{N\,\rm OR} &\simeq& \frac{1}{4\varepsilon_{N}\sigma A_{\rm OR}T_{0}^{3}}, \label{eq.23a.pep} \\
 C_{\rm OR} &=& m_{\rm OR}c_{\rm OR}  \label{eq.23b.pep},
\end{eqnarray}
where it is important to notice that $\theta_{N\,\rm OR}$ does not depend 
on the resonator's emissivity (assuming $\varepsilon_{N}\ll\varepsilon_{\rm OR}$ and that 
$r_{\rm OR}\simeq r_{N}$, where $r_{\rm OR}$ is a representative radius 
of the optical resonator), $\varepsilon_{\rm OR}$, and for this 
reason it will usually be similar to $\theta$.  If such condition is met 
and the thermal capacitance of the shield and the resonator are also similar, 
Eq.\ (\ref{eq.10.pep}) with an extra layer can be used instead of Eq.\ (\ref{eq.22}). 

\subsection{Conductive heat transfer \label{sec.3.2}}
Thermal shield layers need, of course, mechanical supports between them. In this section, the effect 
of such a support structure on the performance of the thermal shields (2 layers) is analyzed. 
The supports or thermal links are supposed to have a very low thermal conductivity 
and, therefore, the lumped model approximation is not adequate. It is not 
guaranteed that the temperature is uniform along the supports~\cite{Incropera,Nofrarias,0264-9381-23-17-005}. 
Consequently, the spatial gradient term of the Fourier heat transfer equation needs to be included. 
In this case, the transfer function of a shield is ---see Appendix~B for details: 
\begin{equation}\label{eq.10b.pep}
 \widetilde{H}(\omega) = \frac{{\rm sinh}\ell q_{\rm s} + q_{\rm s}\theta\kappa_{\rm s} A_{\rm s}}{(1 + i \omega\tau)
 {\rm sinh}\ell q_{\rm s} + q_{\rm s} \theta \kappa_{\rm s} A_{\rm s} {\rm cosh}\ell q_{\rm s}}
\end{equation}
with 
\begin{equation}
 q_{\rm s}^{2} \equiv \frac{\rho_{s}c_{s}}{\kappa_{\rm s}} i\omega,
\end{equation}
where $\kappa_{\rm s}$, $c_{\rm s}$, $\rho_{\rm s}$, $A_{\rm s}$ and $\ell$ are the 
conductivity, the specific heat, the density, the cross section and the length of the supports. 
Note that the supports are assumed to have the same length and the cross-section is 
the sum of all of them.
For the derivation of the transfer function it has been assumed that 
no radiative heat transfer occurs between the supports and the thermal shields. Therefore, the 
model is accurate if the supports are covered with high reflective coating.
In Sec.~\ref{sec.4.2}, a detailed numerical analysis is shown and in Sec.~\ref{sec.5} they are compared with 
FEM simulations.

\section{Concentric spheres and concentric cylinders thermal shields \label{sec.4}}
\begin{figure}[h!]
 \begin{center}
  \includegraphics[width=0.8\linewidth]{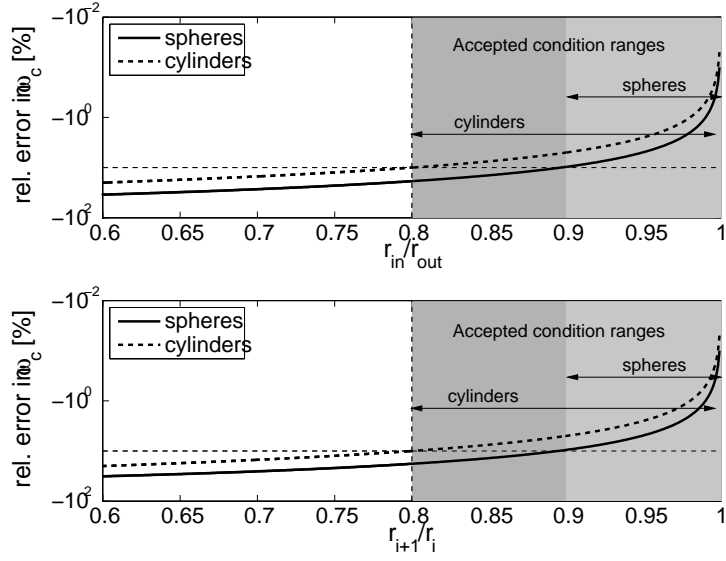}
 \end{center}
\caption{Top: relative error in $\omega_{\rm c}$ due to approximating 
the volume of a layer to $4\pi r^{2}h$ (or $2\pi rh\ell$ for cylinders). 
Bottom: relative error in $\omega_{\rm c}$ due to the assumption that 
the distance between two consecutive layers is 
zero ($r_{i}=r_{j}$) when it is not.  \label{fig.4}} 
\end{figure}

Basically two geometries are used as thermal shields: concentric hollow spheres 
(or cubes) and concentric hollow cylinders (or rectangular cuboids). The cut-off 
angular frequency $\omega_{\rm c}$ of a thermal shield defines completely 
the attenuation of temperature fluctuations if we consider solely radiation. 
It is given in Eq.~(\ref{eq.7.pep}),  
where $\beta$ depends on the geometry of the shields and the emissivity of the 
material~\cite{Incropera}:
 \begin{eqnarray}
 \beta_{ij}&=\frac{1}{\varepsilon_{j}} + 
	\frac{1-\varepsilon_{i}}{\varepsilon_{i}}\left(\frac{r_{j}}{r_{i}}\right)^2, & {\rm for \ spheres} \label{eq.15a.pep} \\
 \beta_{ij}&=\frac{1}{\varepsilon_{j}} +  \frac{1-\varepsilon_{i}}{\varepsilon_{i}}\left(\frac{r_{j}}{r_{i}}\right), & 
 {\rm for \ (infinitely \ long) \ cylinders} \nonumber \\ \label{eq.15b.pep}
 \end{eqnarray}
where $r$ is the radius. The view factors for finite cylinders, cubes and cuboids are 
more complicated than the ones given in Eqs.~(\ref{eq.15a.pep}) and (\ref{eq.15b.pep}), 
however, the approximation of cubes to spheres and finite cylinders and cuboids to 
infinite cylinders yields accurate results ---see Sec.~\ref{sec.4.2}.
If the shields are assumed to have approximately equal size ($r_{i}\approx r_{j}$) 
and high reflectivity materials are used ($\varepsilon\ll 1$), 
Eqs.~(\ref{eq.15a.pep}) and (\ref{eq.15b.pep}) reduce to $\beta\simeq2/\varepsilon$ and 
the cut-off angular frequency of a thermal shield is 
\begin{equation}\label{eq.16.pep}
 \omega_{\rm c} = \frac{2\varepsilon \sigma T_{0}^3}{\rho c h}, 
 \end{equation}
where $h$ is the thickness of the layers. In Sec.~\ref{sec.3.1}, we 
assumed that $\tau$ is the same for each shield.
Equation~(\ref{eq.16.pep}) shows that 
this assumption is valid for the aforementioned conditions.
 However, in Eq.\ (\ref{eq.16.pep}) two simplifications have been used: 
(i) the volume of the hollow sphere (or cylinder) has been approximated to 
$4\pi r^{2}h$ (and similarly for the cylinder) 
and, (ii) the radius 
of all layers has been considered the same ($r_{i}=r_{j}$).
Both of these assumptions cause errors in the calculated cut-off 
angular frequency. To keep them below 10\% ---see Fig.~\ref{fig.2b}, 
the ratio between the inner and outer radii of the layer $i$ has to be 
$r_{i, {\rm in}}/r_{i, {\rm out}}\geq$0.9 for spheres (the layer thickness has to 
be $h\leq0.1r_{\rm out}$) and $r_{i, {\rm in}}/r_{i, {\rm out}}\geq$0.8 for cylinders ---see Fig.\ \ref{fig.4} (top); and 
the ratio between the radius of two consecutive layers has to be
$r_{i+1}/r_{i}\geq$0.9 for spheres and $r_{i+1}/r_{i}\geq$0.8 for cylinders ---see Fig.\ \ref{fig.4} (bottom). 
Both conditions, thin layers and small distance between the layers, 
are usually met when designing and constructing thermal shields.

The only assumption in Sec.~\ref{sec.3.1}, which has not yet been 
justified, is the one concerning the temperature 
homogenization of the thermal shields. This matters, when a point-like heat source 
is present. A homogeneous temperature distribution can be assumed if 
the time constant of the shield itself is much smaller than the time constant 
of the radiative heat transfer, i.e., $\tau\gg\tau_{\rm shield}$.
The time constant of the shield is (for the spherical case)
\begin{equation}
 \tau_{\rm shield}=\theta_{\rm shield}C_{\rm shield}\sim\frac{2\pi r^2\rho c}{\kappa}
\end{equation}
and considering thin layers ($h=0.1r$):
\begin{equation}
 \frac{\tau_{\rm shield}}{\tau}=\frac{4\pi r}{0.1\kappa}\varepsilon\sigma T_{0}^3,
\end{equation}
which is much smaller than one for high reflectivity and high conductivity materials. 

\subsection{Selection of the material}
Given a requirement for the attenuation of temperature fluctuations, 
the thermal shields can be optimized in either mass 
or volume. In this section, this is briefly analyzed. For two thermal 
shields of different materials, A and B, having 
the same cut-off angular frequency the ratio between the thickness 
of the shields is (assuming both materials have the same 
emissivity, which can be easily done by coating the material)
 \begin{equation}\label{eq.19.pep}
  \frac{h_{\rm B}}{h_{\rm A}}=\frac{\rho_{\rm A}c_{\rm A}}{\rho_{\rm B}c_{\rm B}}
 \end{equation}
and the mass ratio is
\begin{equation}\label{eq.20.pep}
 \frac{m_{\rm A}}{m_{\rm B}}=\frac{r^{2}_{\rm A}c_{\rm B} }{r^{2}_{\rm B}c_{\rm A}}.
\end{equation}
If we want to minimize the volume for a fixed mass of the 
layer ($m_{\rm A}= m_{\rm B}$), 
a material with a low specific heat should be chosen following 
Eq.\ (\ref{eq.20.pep}). In contrast, if 
the mass should be minimized for a given radius of the 
layer ($r_{\rm A}=r_{\rm B}$), a material with a high specific 
heat should be selected. 
 
\subsection{Numerical analysis \label{sec.4.2}}
\begin{figure}[h!]
 \begin{center}
 \includegraphics[width=0.7\linewidth]{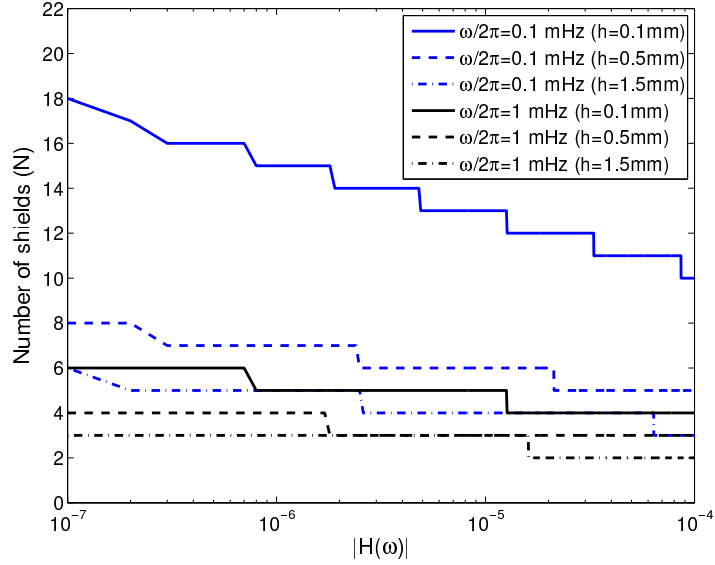}
 \end{center}
\caption{Number of shields needed as a function of the required 
attenuation for 0.1\,mHz and 1\,mHz and 
different shield thickness, $h$.  \label{fig.5}}
\end{figure}

In this section, a numerical analysis of the equations shown in 
Sec.~\ref{sec.3.1} and Sec.~\ref{sec.3.2} is performed.
First, Eq.~(\ref{eq.10.pep}) is used to calculate the required 
number of shields, $N$, for a given temperature stability 
requirement. If the attenuation is needed for $\omega\geq10\omega_{\rm c}$, 
the shields can be assumed uncoupled ---see Fig.~\ref{fig.2}, and $N$ is:
\begin{equation}
  N \geq \Bigg\lceil- \frac{\log|\widetilde{H}_{\rm req}|}{\log{\omega/\omega_{\rm c}}}\Bigg\rceil \ {\rm for} \ \omega\gtrsim10\omega_{\rm c},
\end{equation}
where $\lceil \cdot \rceil$ is the ceiling function. Figure~\ref{fig.5} shows the required 
number of shields for different attenuation levels: from $10^{-7}$ to $10^{-4}$ 
and for two frequencies: $\omega/2\pi$=0.1\,mHz and 1\,mHz. The shields are 
assumed to be made of gold coated aluminum ($\varepsilon$=0.03) 
with a thickness of 0.1\,mm, 0.5\,mm and 1.5\,mm, which correspond to 
cut-off angular frequencies of 60\,$\mu$Hz, 
12\,$\mu$Hz and 4\,$\mu$Hz, respectively. 

\begin{figure}[h!]
 \begin{center}
 \includegraphics[width=0.8\linewidth]{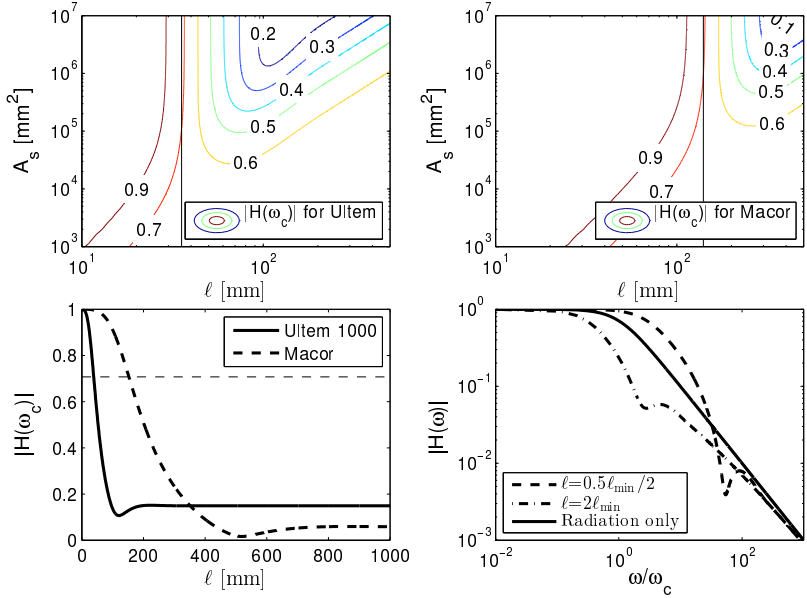}
  \end{center}
\caption{Transfer function for different support lengths and cross-sections 
for two materials and $h$=0.5\,mm. Top left: Ultem 1000. Top right: Macor. 
The vertical dashed lines indicate $\ell_{\rm min}$.
Bottom left: the attenuation of temperature fluctuations at 
$\omega_{\rm c}$ for different support lengths. We set in this 
calculations $r_{i+1}/r_{i}=0.9$ and $A_{\rm s}/A_{j}=0.1$.
If $\ell<\ell_{\rm min}$ the attenuation is degraded 
around $\omega_{\rm c}$ compared to a system without the supports 
(the attenuation at $\omega_{\rm c}$ is $1/\sqrt{2}$  ---horizontal 
dashed line). Once $\ell>\ell_{\rm min}$, increasing the length of the 
supports improves the damping of temperature fluctuations. This 
improvement reaches an asymptotic value not too far from $\ell_{\rm min}$.
Bottom right: transfer functions as a function of the frequency 
for two scenarios with different support lengths and the radiative case only. 
\label{fig.6}}
\end{figure}

Next, we design the supports between the layers such that they do not 
degrade the attenuation of the temperature fluctuations. To do so, we 
employ Eq.\ (\ref{eq.10b.pep}). Figure~\ref{fig.6} (top) shows the 
attenuation at the radiative cut-off angular frequency ---cf.\ Eq.\ (\ref{eq.16.pep}), 
for two materials as a function of the length of the supports $\ell$ 
and the total cross section $A_{\rm s}$ of the supports. 
For the calculations, the layer thickness has been set to $h$=0.5\,mm 
and as material we chose aluminum. The distance between the layers $\ell$ 
is kept such that $r_{i+1}/r_{i}=0.9$ (the 
distance between the layers is $0.1r_{i}$). The total cross-section of 
the supports has been constraint to $A_{j}/10$ (the supports only 
cover 10\% of the thermal shields area). For larger areas the model 
is not accurate since $\beta$ is no longer 
of the form given in Eq.\ (\ref{eq.15a.pep}) ---see Sec.~\ref{sec.5}.
However, such large areas for the supports are unusual for typical designs 
of thermal shields. The properties of the support materials are given 
in Table~\ref{table.2}. The materials are chosen 
due to their low thermal conductivity.
\begin{table}[h!]
\centering
 \caption{Properties of the support materials used for the numerical 
 evaluation in Fig.~\ref{fig.6}. \label{table.2}} 
\begin{tabular}{lccc}
\hline \\
 Material & $\kappa_{\rm s}$ [W\,m$^{-1}$\,K$^{-1}$] & $\rho_{\rm s}$ [kg\,m$^{-3}$] & $c_{\rm s}$ [J\,kg$^{-1}$\,K$^{-1}$] \\
 \hline
 Ultem 1000 & 0.122 & 1280 & 2000 \\
 Macor & 1.46 & 2520 & 790 \\
 \hline
\end{tabular}
\end{table}

Figure~\ref{fig.6} shows how the behavior of the system changes at a certain length $\ell_{\rm min}$: 
For fixed $\ell \ (>\ell_{\rm min})$, increasing the cross-section of the supports 
improves the attenuation of temperature fluctuations and for $\ell<\ell_{\rm min}$ diminishes it. This 
turning point, $\ell_{\rm min}$, occurs    
for $\tau_{\rm s} \gtrsim 2\tau$, where
\begin{equation}\label{eq.21}
 \tau_{\rm s} \simeq \frac{\ell^{2}\rho_{\rm s} c_{\rm s}}{\kappa_{\rm s}}
\end{equation}
and $\tau=\theta C$. Therefore, the respective minimum length of the supports is:
\begin{equation}
 \ell_{\rm min} \simeq \left(\frac{\rho c h}{\varepsilon \sigma T_{0}^3} \frac{\kappa_{\rm s}}{c_{\rm s}\rho_{\rm s}  }\right)^{1/2},
\end{equation}
which depends only on the materials of the shields and supports as well as the 
thickness of the layers $h$. For instance, for $h$=0.5\,mm we have 
$\ell_{\rm min}=35.5\,$mm for Ultem 1000 and $\ell_{\rm min}=150$\,mm for Macor. 
It is important to remark that once $\ell>\ell_{\rm min}$ the support 
cross-sections do not have any constraints provided 
$A_{\rm s}/A_{j}<1/10$, which we assumed in the derivation.
Note that it is not necessary that the shield layers are at 
that distance $\ell_{\rm min}$, because the thermal supports can be 
routed conveniently between the shields allowing still a compact designs 
as long as the radiative heat transport between the supports and the layers 
is negligible. The latter assumption of our model can be met by using 
supports with high reflectivity coatings. This at first glance counter-intuitive 
change in behavior occurs since for a certain length increasing the area 
of the support causes an increase of the thermal capacitance, which compensates 
the reduction of the thermal resistance due to the larger cross-section of the support. 

The bottom left plot in Fig.\ \ref{fig.6} shows the attenuation 
at the radiative cut-off angular frequency $\omega_{\rm c}$ for a given $A_{\rm s}$ 
as a function of the length of the supports for Ultem 1000 and Macor. 
The minimum length corresponds to that where the attenuation is $1/\sqrt{2}$, i.e., 
the length where the cut-off angular frequency of the thermal shield 
with supports does not change with respect to the one without supports. The 
minimum length for Ultem is much shorter than the one for Macor and also 
the attenuation drops much faster when increasing the length.
However, in the asymptotic limit the attenuation for Macor is stronger than the Ultem one:
\begin{equation}
  \lim_{\ell \to \infty}H(\omega_{\rm c})=\frac{1}{1 + 
 i + (\frac{i}{\omega_{\rm c}}\rho_{\rm s} c_{\rm s} \kappa_{\rm s})^{1/2}\frac{1}{C_{j}}A_{\rm s}}.
\end{equation}

The right bottom plot shows the transfer function for three scenarios:
(i) $\ell=\ell_{\rm min}/2$, (ii) $\ell=2\ell_{\rm min}$ and (iii) radiation only.
The rest of the parameters are kept the same. The change of behavior in 
the attenuation is again clear. It is also interesting that combining 
the thermal shields and the supports (radiation and conduction) adequately 
a dip in the transfer function can be obtained. Such behavior could be used to design 
thermal shields, where temperature fluctuations have to be attenuated 
in a very narrow frequency band.

The results shown in Fig.\ \ref{fig.6} are for one 
thermal shield consisting of two layers. When nesting $N$ shields the 
transfer function including the supports can be approximated by:
\begin{equation}
 \widetilde{H}(\omega)\simeq \left[\frac{{\rm sinh}\ell q_{\rm s} + q_{\rm s}\theta\kappa_{\rm s} A_{\rm s}}{(1 + i\tau \omega)
 {\rm sinh}\ell q_{\rm s} + q_{\rm s} \theta \kappa_{\rm s} A_{\rm s} {\rm cosh}\ell q_{\rm s}}\right]^{N}.
\end{equation}
This assumption of uncoupled layers yields errors similar to those shown in Fig.~\ref{fig.2}. 
The derivation of the solution for the coupled case is given in Appendix~B.

\section{Analytical solutions and FEM simulations results\label{sec.5}}
\begin{figure}[h!]
 \begin{center}
 \includegraphics[width=1\linewidth]{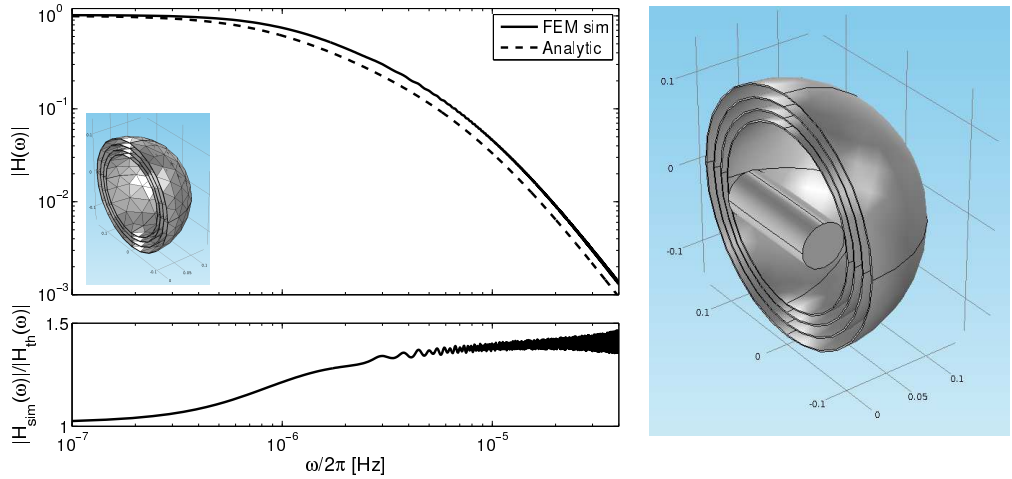}
  \end{center}	
\caption{Left: FEM simulation and theoretical results ---cf.\ Eq.\ (\ref{eq.10.pep}) for 
the transfer function of a thermal shield consisting of four aluminum 
layers ($N$=3) considering only radiation. The bottom panel 
shows the ratio between the results. Right: thermal shield model used for the calculations and 
simulations. The OR (12.5\,cm) is not included in the simulations. 
The mesh is shown in the left figure inset. \label{fig.7}}
\end{figure}

In this section, we compare the results and properties found in the previous section 
with the results of FEM simulations performed with commercial software (Comsol). Figure~\ref{fig.7} 
shows the transfer function, if we take only radiation into account. The 
model considered here consists of four aluminum layers ($N$=3) with $\varepsilon$=0.03 and $h$=1.5\,mm.
The radii of the shields are from the inner shield to the outer: 100\,mm, 111.1\,mm, 123.9\,mm 
and 137.9\,mm, respectively i.e., the ratio $r_{i+1}/r_{i}$ has been kept at approximately 0.9. 
The maximum mesh element size is 31\,mm and the minimum one is 5.5\,mm. The mesh settings 
are valid for the rest of the simulations with finer mesh when including the conductive thermal links.
We use a sweep sine with frequencies from 1\,$\mu$Hz to 
50\,$\mu$Hz as the boundary condition of the outermost layer of the thermal shield. The transfer 
function is calculated as the ratio of the Fourier transformed temperature at the innermost 
and the outermost layer of thermal shield. Figure~\ref{fig.7} shows that the numerical simulations are in very good 
agreement with the analytical results calculated using Eq.\ (\ref{eq.10.pep}). The discrepancy 
between the simulations and the model is close to the expected one due to the fact of 
assuming $\tau_{i}=\tau_{i+1}$ when $r_{i+1}/r_{i}$=0.9 ---see Fig.~\ref{fig.2b} and Fig.~\ref{fig.4}.

\begin{figure}[h!]
 \begin{center}
  \includegraphics[width=0.8\linewidth]{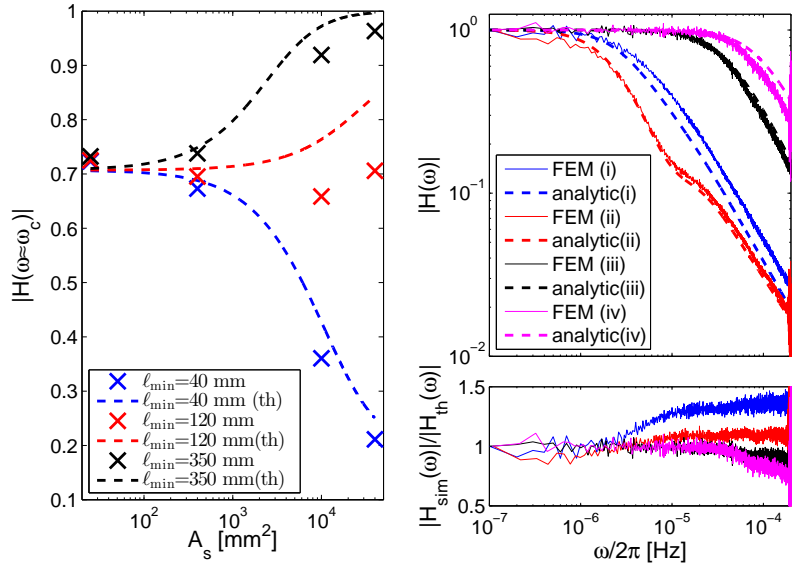}
  \end{center}
\caption{FEM simulations and analytical results. Left: temperature fluctuations 
attenuation at $\omega_{\rm c}$ for $\ell>\ell_{\rm min}$ (blue) 
and $\ell<\ell_{\rm min}$. (red and black). $\ell=$70\,mm for all the simulations. 
Right: transfer functions for different models ---see Table~\ref{table.3}, and 
the ratios between the simulations and the analytical transfer functions, 
which indicate the good agreement between them. The simulations 
failed for frequencies higher than 100\,$\mu$Hz due to numerical 
resolution issues. \label{fig.8}} 
\end{figure}

Figure~\ref{fig.8} (left panel) shows the simulations performed to evaluate the 
behavior found in Sec.~\ref{sec.4.2}, when conductive elements are included in the model. Different 
configurations have been simulated with supports (70\,mm in length) used to connect the thermal shield layers.
The properties of the shields and the supports have been tweaked to achieve different $\ell_{\rm min}$ values.
The simulations were carried out for different cross-sections with an oscillating temperature at $\sim \omega_{\rm c}$ as the boundary condition in the 
outer aluminum layer. Three configurations have been simulated for a given $\ell$=70\,mm and different $\ell_{\rm min}$: 
(i) $\ell_{\rm min}=1.75\ell$=40\,mm (blue trace), (ii) $\ell_{\rm min}=0.58\ell$=120\,mm (red trace) and (iii) 
$\ell_{\rm min}=0.2\ell=350$\,mm. The results from the simulations (``$\times$`` marks) and from the analytical equations (dashed lines) are also in 
good agreement. The right panel in Fig.\ \ref{fig.8} shows the simulated 
and analytical transfer functions for different parameters, when the 
supports are included. The model considers only one shield and the properties have been chosen to simulate 
different expected behaviors. They are summarized in Table~\ref{table.3}. Cubes have been used instead of 
spheres and, therefore, $A_{j}$ needs to be calculated accordingly since it appears in $\theta$ in Eq.\ (\ref{eq.10b.pep}).
The results from the simulations and the analytical calculations are in very well agreement, too.
\begin{table}[h!]
\centering
\caption{
Properties of the supports used in Fig.\ \ref{fig.8} (right panel): $\rho_{\rm s}$=1200\,kg\,m$^{3}$, $c_{\rm s}$=7200\,J\,kg$^{-1}$\,K$^{-1}$, 
$\ell=0.053$\,m; properties of the layers of the thermal shield: aluminum with $h$=1.5\,mm, $\varepsilon$=0.03, $A_{2}$=0.06\,m$^{2}$,
 $r_{2}$=0.1\,m and $r_{1}$=0.153\,m ($2r$ is the length of cube's edge). \label{table.3}}
\begin{tabular}{lll}
\hline 
Model & $\kappa_{\rm s}$\,[W\,(m\,K)$^{-1}$] & $A_{\rm s}$\,[m$^{2}$]  \\
 \hline
 (i)   & 0.122 & $10^{-3}$  \\
 (ii)  & 0.122 & $6.2\times10^{-3}$ \\
 (iii) & 12.2 & $10^{-3}$ \\
 (iv)  & 12.2 & $6.2\times10^{-3}$ \\
\hline
 \end{tabular}
\end{table}

\begin{figure}[h!]
 \begin{center}
  \includegraphics[width=0.8\linewidth]{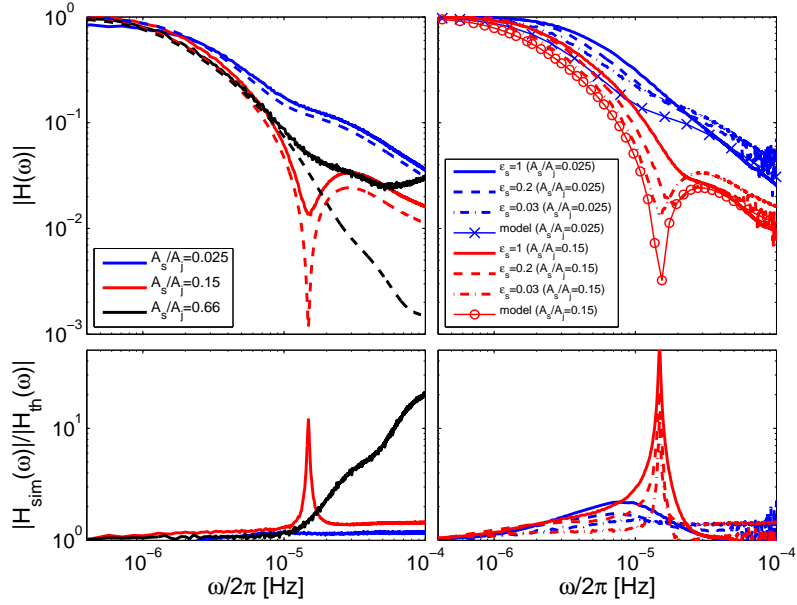}
  \end{center}
\caption{Left: FEM simulations (solid lines) and theoretical transfer functions (dashed lines) 
for different values of the supports cross sections and no radiation between 
the supports and the shields, $\varepsilon_{\rm s}$=0. The model agrees with the simulations
when the total area of the supports is significantly smaller than the area of the shield ($A_{\rm
s}/A_{j}\lesssim0.15$) as previously stated. The bottom plot shows the ratio between the 
simulations and the analytical transfer functions. The discrepancies are significant for 
$A_{\rm s}/A_{j}$=0.66 at high frequencies. Right: FEM simulations (solid and dashed lines) for different 
supports emissivities and $A_{\rm s}/A_{j}$ ratios and the theoretical transfer functions 
(dash-dotted lines) where $\varepsilon_{\rm s}=0$. The difference between the simulations 
and the model appears for large emissivity values. The bottom plot shows the 
ratios between the numerical simulations transfer functions and the analytical model. 
\label{fig.9}}
\end{figure}

Finally, simulations have been performed to quantify two limitations of the
analytical model and its assumptions: (i) the ratio of the cross-section of the
supports and the area of the shields was assumed to be sufficiently small ---see
Sec.~\ref{sec.4.2} and, (ii) radiative heat transfer between the supports and the
shields was neglected. If (i) is not met, the view factors used for the derivation in Sec.\ref{sec.4.2}
are not valid anymore. Figure~\ref{fig.9} (left) shows the comparison between the
simulations (solid lines) and the model (dashed lines). The results are in good
agreement for ratios smaller than $\sim$0.15. The disrepancy at the dip 
for $A_{\rm s}/A_{j}=0.15$ is mainly due to the limited frequency resolution of the estimated 
transfer function from the simulations.
For larger ratios (black trace), the errors become significant. For instance, for $A_{\rm s}/A_{j}=0.66$ the
theoretical transfer function and the one obtained by FEM simulations differ by a factor of
$\sim10$ at high frequencies. However, such large ratios are rarely used in thermal shield
designs. The right panel in Fig.\ \ref{fig.9} compares the simulations when including
radiative heat transfer between the supports and the shield layers. 
The simulations (solid and dashed lines) and the model (dash-dotted lines) are in 
well agreement for small $\varepsilon_{\rm s}$.

\section{Conclusion \label{conclusion}}
Highly sensitive interferometers and experiments testing fundamental physics require an isolation 
against external and environmental influences. In particular, the attenuation of external temperature 
fluctuations by several orders of magnitude at low frequencies (sub-milli-Hertz and milli-Hertz) using thermal shields is crucial. 
Such thermal shields are typically designed by numerical FEM simulations, 
which are time consuming and precision limited. Otherwise, simple analytical ''toy``
 models that do not reflect all the features of the system are employed. However, an accurate 
analytical method is to our knowledge still lacking. We derived in this paper a fast and 
accurate method to calculate the performance of thermal shields in the frequency domain. This analysis has also 
yielded interesting results, such as that for some configurations a dip in the transfer function can be 
achieved in a narrow frequency range, and that the cross-section of the supports 
can be increased without jeopardizing the shields performance as far as they reach a critical minimum length. 
The former property is of interest for experiments with a narrow and well-known frequency range, such as BOOST. 
The latter can be useful for space experiments to design mechanically stable shields without loosing thermal insulating capability.
The analytical model has been validated by comparing it with the results obtained by FEM simulations. 
This method allows for quick investigations of the design of thermal shields to optimize its performance, 
which later can be assessed and fine-tuned by FEM simulations.

\appendix

\section*{Appendix A: Derivation of the transfer function considering only radiation \label{app.a}}
\setcounter{equation}{0}
\renewcommand{\theequation}{A{\arabic{equation}}}

\begin{figure}[h!]
 \begin{center}
 \includegraphics[width=0.65\linewidth]{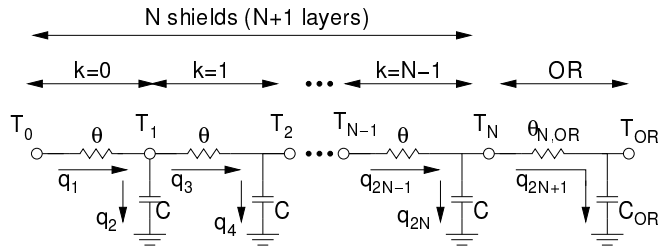}
 \end{center}
\caption{Electrical circuit equivalent to a thermal shield consisting of $N$ shields
including the optical resonator. The thermal resistance and heat capacitance 
are assumed to be the same for all of the layers except the resonator one.}\label{fig.app.1} 
\end{figure}

In this appendix, we derive the transfer function considering only radiation. 
The electrical circuit equivalent to $N$ thermal shields and the optical resonator 
is shown in Fig.\ \ref{fig.app.1} ---see Sec.~\ref{sec.3.1} for the definitions of $\theta_{ij}$ and $C_i$.
Note that this analogy is only valid as far as the linearization used in Eq.\ \ref{eq.5.pep} of the 
radiative thermal heat transfer holds.

The transfer function is defined as the ratio between the first and the last 
temperature node, i.e., $\widetilde{T}_{\rm OR}/\widetilde{T}_{0}$, where \emph{tildes} ($\ \widetilde{}$\ ) stand 
for Fourier transforms.
The transfer function is found by solving the following system of equations, which is constructed by 
applying the Kirchhoff's circuit laws.
\begin{eqnarray}
 \widetilde{T}_{k}-\widetilde{T}_{k+1}-\theta q_{2k+1} &=& 0, \ k=0...N-1  \label{eq.ap.1} \\
 \widetilde{T}_{k+1}-\frac{1}{s C} q_{2k+2} &=&0, \ k=0...N-1 \\ 
 q_{2k+1} - q_{2k+2} - q_{2k+3} &=&0, \ k=0...N-1 \\
 \widetilde{T}_{N}-\widetilde{T}_{\rm OR} - \theta_{\rm OR}q_{2N+1}&=&0 \\
 \widetilde{T}_{\rm OR} - \frac{1}{sC_{\rm OR}}q_{2N+1}&=&0, \label{eq.ap.2}
\end{eqnarray}
where $q$ is the heat flow (the current in the analogy of electrical circuits) from one node to 
the other in units of Watt and the temperature (the voltage in the analogy), and $s=i\omega$ is the Laplace 
variable with the Fourier frequency $\omega$. All the thermal resistances and capacitances are assumed 
to be equal (except the optical resonator's one). The solution of this system of equations can be written 
in a compact form using the Pascal triangle numbers. If one does not consider the 
optical resonator the transfer function is the one given by Eq.\ (\ref{eq.10.pep}). The one including 
the resonator is given by Eq.\ (\ref{eq.22}).

\section*{Appendix B: Derivation of the transfer function including conductive links \label{app.b}}
\setcounter{equation}{0}
\renewcommand{\theequation}{B{\arabic{equation}}}

\begin{figure}[h!]
 \begin{center}
  \includegraphics[width=0.25\linewidth]{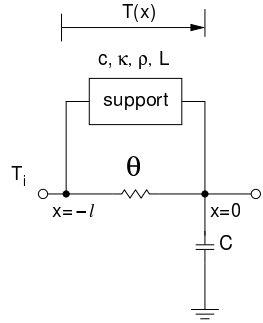}
 \end{center}
\caption{Model including the conductive link in a thermal shield layer. Typically, the support 
cannot be modeled as a thermal resistance since large thermal gradients are 
present along it. Instead the Fourier heat transfer equation needs to be solved. The model does 
not include radiative heat transfer from the supports to the shields.
\label{fig.2c}}
\end{figure}

As stated in Sec.~\ref{sec.3.2} the lumped model is not adequate to represent 
the conductive heat transfer via the supports connecting the shields. Consequently, 
the Fourier heat equation has to be used. Radiative heat transfer is not considered between 
the supports and the thermal shields, i.e., the supports are assumed to be coated 
with materials with very low emissivity or, alternatively they have small areas of sight.
In this scenario, the conductive heat 
transfer within the supports occurs only in one direction ---see Fig.\ \ref{fig.2c}:
\begin{equation}\label{eq.app.0}
 \rho_{\rm s} c_{\rm s} \frac{\partial T(x,t)}{\partial t} = \kappa_{\rm s} \frac{\partial^{2} T(x,t)}{\partial x^{2}}, \quad -\ell\leq x\leq 0,
\end{equation}
where $\rho_{\rm s}$, $c_{\rm s}$ and $\kappa_{\rm s}$ are the density, the specific heat and the thermal conductivity of the
supports. The transfer function is found by applying the Fourier transform to Eq.~(\ref{eq.app.0})~\cite{Carslaw}:
\begin{equation}\label{eq.app.1}
 \frac{d^2\widetilde{T}(x)}{dx^2}-q_s^2\widetilde{T}(x)=0,
\end{equation}
where $q_{\rm s} = \left(i\frac{\rho_{s}c_{s}}{\kappa_{\rm s}}  \omega \right)^{1/2}$. Equation~(\ref{eq.app.1}) 
is solved with the following boundary conditions (known boundary temperature at the outer most layer, $\widetilde{T}_{0}$, 
and continuous heat flux at the interface):
\begin{eqnarray}
\widetilde{T}(-\ell)&=&\widetilde{T}_{0} \\
-\kappa_{\rm s}A_{\rm s}\frac{d\widetilde{T}(x)}{dx}\Bigg|_{0}&=&i \omega C\widetilde{T}(0) + \frac{\widetilde{T}(0)-\widetilde{T}(-\ell)}{\theta},
\end{eqnarray}
The transfer function is $\widetilde{T}(0)/\widetilde{T}_{0}$ and the 
solution is given in Eq.~(\ref{eq.10b.pep}).

\begin{figure}[h!]
 \begin{center}
  \includegraphics[width=0.8\linewidth]{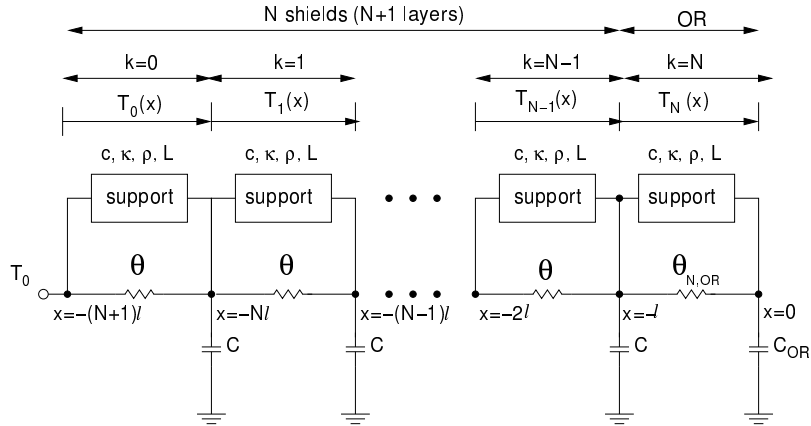}
 \end{center}
\caption{Model including the conductive link and the optical 
resonator for $N$ thermal shields and the OR.
\label{fig.app.last}}
\end{figure}

The exact solution for $N$ shields plus the resonator and the supports is found by solving the following 
system of differential equations ---see Fig.~\ref{fig.app.last}:
\begin{eqnarray}
  \frac{d^2\widetilde{T}_{k}(x)}{dx^2}-q_{\rm s}^2\widetilde{T}_{k}(x)=0, \quad -(N-(k-1))\ell\leq &x& \leq -(N-k)\ell
 \end{eqnarray}
with $k=0\ldots N$. The boundary conditions are that the temperature and the heat flux are continuous 
across the interfaces:
\begin{eqnarray}
 \widetilde{T}_{0}(-(N+1)\ell)&=&\widetilde{T}_{0} \\
 \widetilde{T}_{k}(-(N-k)\ell)&=&\widetilde{T}_{k+1}(-(N-k)\ell)\\
 -\kappa_{\rm s}A_{\rm s}\frac{d\widetilde{T_{k}}(x)}{dx}\Bigg|_{-(N-k)\ell}&=&i \omega C\widetilde{T_{k}}(-(N-k)\ell)+ \nonumber \\
  &+& \frac{\widetilde{T_{k}}(-(N-k)\ell) - \widetilde{T_{k}}(-(N-k+1)\ell)}{\theta}  \\
  -\kappa_{\rm s}A_{\rm s}\frac{d\widetilde{T_{N}}(x)}{dx}\Bigg|_{0}&=&i \omega C_{\rm OR}\widetilde{T_{N}}(0) 
  + \frac{\widetilde{T_{N}}(0) -\widetilde{T_{N}}(-\ell)}{\theta_{\rm OR}},
\end{eqnarray}
where here $k=0\ldots N-1$.

\section*{Acknowledgments}
The authors appreciate valuable discussions with Alexander Milke, 
Thilo Schuldt and Martin Siemer. This work was supported by the German space agency 
(Deutsches Zentrum f\"ur Luft- und Raumfahrt DLR) with 
funds provided by the Federal Ministry of Economics and Technology under grant numbers 50 QT 1401 and the DFG Sonderforschungsbereich (SFB) 1128 Relativistic Geodesy and Gravimetry with Quantum Sensors (geo-Q).
. 


\begin{thebibliography}{99}

\bibitem{0264-9381-29-12-124016} P. Amaro-Seoane, S. Aoudia, S. Babak, P. Bin\'etruy, E. Berti, A. Boh\'e, C.
Caprini, M. Colpi, N. J. Cornish, K. Danzmann, J.-F. Dufaux, J. Gair, O. Jennrich, P. Jetzer, A. Klein, R. N. Lang, A. Lobo, T. Littenberg, S. T.
McWilliams, G. Nelemans, A. Petiteau, E. K. Porter, B. F. Schutz, A. Sesana, R.
Stebbins, T. Sumner, M. Vallisneri, S. Vitale, M. Volonteri, and H. Ward, 
``Low-frequency gravitational-wave science with eLISA/NGO,'' Class. Quantum Grav. \textbf{29}(12), 124016 (2012).
\bibitem{0264-9381-26-15-153001} O. Jennrich, ``LISA technology and instrumentation,'' Class. Quantum Grav. \textbf{26}(15), 153001 (2009).
\bibitem{Mueller1} G. Mueller, P. McNamara, I. Thorpe, and J. Camp, ``Laser frequency stabilization for LISA,'' NASA GSFC Tech. Report 20060012084 (2005).
\bibitem{heinzel.grace} B. S. Sheard, G. Heinzel, K. Danzmann, D. A. Shaddock, W. M. Klipstein, and W. M Folkner, 
``Intersatellite laser ranging instrument for the GRACE follow-on mission,'' J. Geodesy \textbf{86}(12), 1083-1095 (2012).
\bibitem{GRACEFO} W. Folkner, G. de Vine, W. Klipstein, K. McKenzie, D. Shaddock, 
R. Spero, R. Thompson, D. Wuchenich, N. Yu, M. Stephens, J. Leitch, M. Davis, J. de Cino, C. Pace, and R. Pierce, ``Laser frequency stabilization for GRACE-II,''
Jet Propulsion Laboratory, California Institute of Technology Tech. Report (2010).
\bibitem{PhysRevLett.88.010401} C. Braxmaier, H. M\"uller, O. Pradl, J. Mlynek, A. Peters, and S. Schiller, ``Tests of relativity using a cryogenic optical resonator,''
Phys. Rev. Lett. \textbf{88}, 010401 (2001).
\bibitem{PhysRevD.81.022003} M. E. Tobar, P. Wolf, S. Bize, G. Santarelli, and V. Flambaum, 
``Testing local Lorentz and position invariance and variation of fundamental constants by searching the derivative of the comparison frequency 
between a cryogenic sapphire oscillator and hydrogen maser,'' Phys. Rev. D \textbf{81}, 022003 (2010).
\bibitem{PhysRevLett.91.020401} H. M\"uller, S. Herrmann, C. Braxmaier, S. Schiller, and A. Peters, 
``Modern Michelson-Morley experiment using cryogenic optical resonators,'' Phys. Rev. Lett. \textbf{91}, 020401 (2003).
\bibitem{PhysRevD.80.105011} S. Herrmann, A. Senger, K. M\"ohle, M. Nagel, E. V. Kovalchuk, and A. and Peters, 
``Rotating optical cavity experiment testing Lorentz invariance at the ${10}^{-17}$ level,'' Phys. Rev. D \textbf{80}, 105011 (2009).
\bibitem{0264-9381-18-13-312} C. L\"ammerzahl, H. Dittus, A. Peters, and S. Schiller,
  ``OPTIS: a satellite-based test of special and general relativity,'' Class. Quantum Grav. \textbf{18}(13), 2499 (2001).
\bibitem{2012arXiv1203.3914L} J. A. Lipa, S. Buchman, S. Saraf, J. Zhou, A. Alfauwaz, J. Conklin, G. D. Cutler, and R. L. Byer, 
``Prospects for an advanced Kennedy-Thorndike experiment in low Earth orbit,'' ArXiv e-prints 1203.3914 gr-qc (2012).
\bibitem{6702260} A. Milke, D. N Aguilera, N. G\"urlebeck, T. Schuldt, S. Herrmann, K. D\"oringshoff, R. Spannagel, 
C. L\"ammerzahl, A. Peters, B. Biering, H. Dittus, and C. Braxmaier, ``A space-based optical Kennedy-Thorndike experiment testing special relativity,'' 
European Frequency and Time Forum International Frequency Control Symposium (EFTF/IFC), 912-914 (2013). 
\bibitem{RevModPhys.83.11} A. Alan Kosteleck\'y and N. Russell, ``Data tables for Lorentz and $CPT$ violation,'' Rev. Mod. Phys. \textbf{83}, 11-31 (2011).
\bibitem{0034-4885-77-6-062901} J. D. Tasson, ``What do we know about Lorentz invariance?,'' Reports on Progress in Physics \textbf{77}(6), 062001 (2014).
\bibitem{PhysRevA.77.033847} S. A. Webster, M. Oxborrow, S. Pugla, J. Millo, and P. Gill, ``Thermal-noise-limited optical cavity,'' Phys. Rev. A \textbf{77}(3), 033847 (2008). 
\bibitem{Argence:12} B. Argence, E. Prevost, T. L\'ev\`eque, R. Le Goff, S. Bize, P. Lemonde, and G. Santarelli, ``Prototype of an ultra-stable optical cavity for space applications,''
Opt. Express \textbf{20}(23), 25409--25420 (2012).
\bibitem{Amaira_2013} S. Amairi, T. Legero, T. Kessler, U. Sterr, J. Wübbena, O. Mandel, and P. Schmidt, 
``Reducing the effect of thermal noise in optical cavities,'' Appl. Phys. B \textbf{113}(2), 233-242 (2013).
\bibitem{:/content/aip/journal/rsi/85/11/10.1063/1.4898334}
   Q-F. Chen, A. Nevsky, M. Cardace, S. Schiller, T. Legero, S. H\"afner, A. Uhde, and U. Sterr,
   ``A compact, robust, and transportable ultra-stable laser with a fractional frequency instability of 1$\times$10−15,''
   Rev. Sci. Instrum. \textbf{85}(11) (2014).
\bibitem{1996SPIE.2857...58E} M. J. Edwards, E. H. Bullock, and D. E. Morton, ``Improved precision of absolute thermal-expansion measurements for ULE glass,''
Advanced Materials for Optical and Precision Structures, Society of Photo-Optical Instrumentation Engineers (SPIE) Conference Series \textbf{2857}, 58-63 (1996).
\bibitem{Birch:88} K. P. Birch and P. T. Wilton, ``Thermal expansion data for Zerodur from 247 to 373 K,'' Appl. Opt. \textbf{27}(14), 2813--2815 (1988).
\bibitem{PhysRevA.77.053809} J. Alnis, A. Matveev, N. Kolachevsky, T. Udem, and T. W. H\"ansch, 
``Subhertz linewidth diode lasers by stabilization to vibrationally and thermally compensated ultralow-expansion glass Fabry-P\'erot cavities,'' Phys. 
Rev. A \textbf{77}(5), 053809 (2008).
\bibitem{0022-3727-28-9-008} M. Notcutt, C. T. Taylor, A. G. Mann, and D. G. Blair, ``Temperature compensation for cryogenic cavity stabilized lasers,''
J. Phys. D: Appl. Phys. \textbf{28}(9), 1807 (1995).
\bibitem{Legero:10} T. Legero, T. Kessler, and U. Sterr, ``Tuning the thermal expansion properties of optical reference cavities with fused silica mirrors,'' 
J. Opt. Soc. Am. B \textbf{27}(5), 914--919 (2010).
\bibitem{:/content/aip/proceeding/aipcp/10.1063/1.2405044} H. Peabody and S. M. Merkowitz, ``Low frequency thermal performance of the LISA sciencecraft,''
\bibitem{PhysRevD.87.102003} M. Nofrarias, F. Gibert, N. Karnesis, A. F. Garc\'ia, M. Hewitson, G. Heinzel, and K. Danzmann,
``Subtraction of temperature induced phase noise in the LISA frequency band,'' Phys. Rev. D \textbf{87}(10), 102003 (2013).
AIP Conference Proceedings \textbf{873}, 204-209 (2006).
\bibitem{Incropera} T. L. Bergman, A. S. Lavine, F. P. Incropera, and D. P. DeWitt, \emph{Fundamentals of Heat and Mass Transfer} (Wiley, 2011).
\bibitem{Nofrarias} M. Nofrarias, ``Thermal diagnostics in the LISA technology package,'' Ph.D Thesis, Universitat de Barcelona (2007).
\bibitem{0264-9381-23-17-005} A. Lobo, M. Nofrarias, J. Ramos-Castro, and J. Sanjuan, ``On-ground tests of the LISA PathFinder thermal diagnostics system,'' 
Class. Quantum Grav. \textbf{23}(17), 5177 (2006).
\bibitem{Carslaw} H. S. Carslaw and J. C. Jaeger, \emph{Heat Conduction in Solids} (Oxford Science Publications, 1986).
\end{thebibliography}
\end{document}